\pdfoutput=1

\documentclass[useAMS,usenatbib]{mn2e}

\usepackage{pifont}
\usepackage[final]{graphicx}
\usepackage{amssymb}
\usepackage{color}

\title[Central gas dynamics in NGC\,1068]{Gas dynamics of the central few parsec region of NGC\,1068 fuelled by the evolving nuclear star cluster}
\author[M. Schartmann et al.]
  {M.~Schartmann$^{1,2,}$\thanks{E-mail: schartmann@mpe.mpg.de},
   A.~Burkert$^{1,2,}$\thanks{Max Planck Fellow},
   M.~Krause$^{1,2}$,
   M.~Camenzind$^{3}$,
   \newauthor
   K.~Meisenheimer$^{4}$
   and R.~I.~Davies$^{1}$\\
$^{1}$Max-Planck-Institut f\"ur extraterrestrische Physik, Giessenbachstra\ss e, D-85748 Garching, Germany\\
$^{2}$Universit\"ats-Sternwarte M\"unchen, Scheinerstra\ss e 1, D-81679 M\"unchen, Germany\\
$^{3}$ZAH, Landessternwarte Heidelberg, K\"onigstuhl 12, D-69117 Heidelberg, Germany\\
$^{4}$Max-Planck-Institut f\"ur Astronomie, K\"onigstuhl 17, D-69117 Heidelberg, Germany}
\begin{document}

\date{Accepted . Received ; in original form }

\pagerange{\pageref{firstpage}--\pageref{lastpage}} \pubyear{2002}

\maketitle

\label{firstpage}

\begin{abstract}
Recently, high resolution observations with the help of the near-infrared
adaptive optics integral field spectrograph SINFONI (Spectrograph for INtegral Field
Observations in the Near Infrared) at the VLT proved the
existence of massive and young nuclear star clusters in the centres of a
sample of Seyfert galaxies. With the help of three-dimensional high resolution hydrodynamical
simulations with the {\sc Pluto} code, we follow the evolution of
such clusters, especially focusing on stellar mass loss feeding gas into 
the ambient interstellar medium and driving turbulence. 
This leads to a vertically wide distributed clumpy or filamentary inflow of gas on
large scales (tens of parsec), whereas a turbulent and very dense disc builds up on
the parsec scale. In order to capture the relevant physics in the inner region, we treat 
this disc separately by viscously evolving the radial surface density
distribution. This enables us to link the tens of parsec scale region
(accessible via SINFONI observations) to the (sub-)parsec scale region 
(observable with the MIDI instrument and via water maser emission). Thereby,
this procedure provides us with an ideal testbed for data comparison. 
In this work, we concentrate on 
the effects of a parametrised turbulent viscosity to generate angular momentum and
mass transfer in the disc and additionally take star formation into account. 
Most of the input parameters are constrained by available observations
of the nearby Seyfert\,2 galaxy NGC\,1068 and we discuss parameter studies for the free 
parameters.
At the current age of its nuclear starburst of 250\,Myr, our simulations yield 
disc sizes of the order of 0.8 to 0.9\,pc, gas masses of $10^6\,M_{\odot}$ and mass
transfer rates of $0.025\,M_{\odot}/\mathrm{yr}$ through the inner rim of the disc. 
This shows that our large scale torus model is able to approximately
account for the disc size as inferred from interferometric observations in the mid-infrared
and compares well to the extent and mass of a rotating disc structure as inferred from water 
maser observations. Several other observational constraints are discussed as well. 
\end{abstract}


\section{Introduction}
\label{sec:introduction}

Often challenged, but on the whole still valid, the {\it Unified Scheme of
Active Galactic Nuclei} \citep{Antonucci_93,Urry_95} constitutes the
standard paradigm for the composition of Active Galactic Nuclei ({\it AGN}). 
Their highly energetic central activity is powered by accretion onto a
supermassive black hole ($10^6-10^{10}\,M_{\odot}$, e.~g.~\citealp{Shankar_04}).
Seeing the central engine or not in this scheme is solely determined by the orientation of a
ring-like dust distribution around the galaxy centre, the so-called {\it
molecular torus}. It constitutes not only the source of obscuration, but also
the gas reservoir for feeding the hot accretion disc. Viewed face-on, most of
the UV-optical continuum emission of the accretion disc (the so-called {\it
big blue bump}) is visible in the
spectral energy distribution (SED), whereas it is blocked by the dust 
in the case of an edge-on view. The absorbed energy is reemitted in the
infrared wavelength regime, giving rise to a pronounced peak in the SED of
many AGN \citep{Sanders_89}. This scenario enables the unification of two
observed classes of galaxies, namely type~1 (direct view onto the central
engine) and type~2 sources (through dust). It was first proposed by \citet{Antonucci_85},
after {\it broad emission lines} -- arising from fast moving gas close to the
central black hole (the so-called broad line region) -- have been detected in
polarized light in the Seyfert~2 galaxy NGC~1068. These photons are scattered
into the line of sight
and polarised by dust and free electrons above the torus opening. In unpolarised 
light, type~2 sources only exhibit narrow emission lines, which originate 
from slower gas motions further away from the black hole, in the so-called
{\it narrow line region}, stretching beyond the opening of the torus funnel. 

Resolving the dusty torus was for the first time possible with the advent of
the mid-infrared interferometer (MIDI) at the VLTI ({\it very large telescope
interferometer}) by combining the light of two of the eight meter class VLT
telescopes. This revealed geometrically thick dust structures on parsec scale 
in a large sample of nearby AGN \citep{Tristram_09}, but with a large variety
of torus properties. Exhaustive observations
of the nearby galaxy NGC~1068 \citep{Jaffe_04,Poncelet_06,Raban_09} and the
Circinus galaxy \citep{Tristram_07} both found the dust to be distributed in a
complex structure. The brightness distribution could be disentangled into two components: a
geometrically thin disc-like structure on sub-parsec to parsec scale and a fluffy or
filamentary torus-like distribution surrounding it. For the case of the Circinus galaxy, the
interferometric signal even showed finestructure. This was interpreted to result
from a clumpy large scale component, which puts severe constraints on theoretical
models. Not being capable of direct imaging, radiative transfer models are
needed for a careful data interpretation and modelling. 

From the theoretical perspective, basically three questions are of current interest:
\begin{dingautolist}{182}
\item What is the structure of tori? 
\item What are the basic shaping processes and how do they sustain the scale height of the torus against gravity?
\item How are tori built up and how do they evolve and feed gas on to the central engine?
\end{dingautolist}

Modelling so far concentrated mainly on several aspects of the first two
questions. A large number of continuous torus models have been used in
radiative transfer calculations in order to compare simulated SEDs with high
resolution observations in the infrared
\citep[e.g.][]{Pier_92,Pier_93,Granato_94,Bemmel_03,Schartmann_05,Fritz_06}, constraining
physical parameters of tori in specific galaxies or whole samples of AGN
tori, like the shape, size, luminosity, etc. 
The most up to date simulations concentrate on clumpiness of the absorbing
dust \citep[e.~g.~][]{Nenkova_02,Hoenig_06,Nenkova_08a,Nenkova_08b,Schartmann_08},
following a suggestion by \citet{Krolik_88}, which helps to prolong the
timescale for the destruction of dust impacted by hot surrounding gas. 
Simultaneous agreement with
high-resolution spectral (e.~g.~NACO or Spitzer) as well as interferometric
data (MIDI) give us a good idea of possible structural properties of these
tori. 

Concerning the second question of the structuring agent and the stability of
the vertical scale height against gravity, several modelling attempts have been put
forward: The torus was claimed to be made up of clumps with supersonic random
velocities, which are sustained by transferring orbital shear energy with the
help of sufficiently elastic collisions between the clumps
\citep[e.~g.~][]{Krolik_88,Beckert_04}. 
Starburst-driven AGN tori were
investigated with the help of hydrodynamical simulations of the interstellar
medium (ISM), concentrating on two effects: (i) a large rate of core-collapse supernova
explosions (following {\it in situ} star formation in the densest region) in a
rotationally supported thin disc \citep{Wada_02,Wada_09} and (ii) feedback due to
strong winds of young and massive stars \citep{Nayakshin_07}.  
Alternative scenarios replace the torus
itself by a hydromagnetic disc wind, in which dusty clouds are embedded and
produce the necessary obscuration differences between type~1 and type~2
objects \citep{Koenigl_94,Elitzur_06} or by nuclear gas discs, which are
warped due to an axisymmetric gravitational potential generated by discs of
young stars \citep[e.g.][]{Nayakshin_05}. \citet{Pier_92}  
claim the importance of infrared radiation pressure in the sub-parsec to
parsec scale part of the torus or nuclear disc component. 
This scenario was elaborated in \citet{Krolik_07}. With the help of analytical
calculations, they find good comparison of gas column densities to
observations for reasonable intrinsic AGN luminosities. 
For a more detailed discussion of available models see
also \citet{Krolik_07}.  

Despite these great successes in pinning down several aspects of physical processes in
the vicinity\footnote{In the following, we mean the parsec scale region around
the black hole, when we talk about the vicinity or the environment of the
black hole!} of black holes, a conclusive global physical model for the build-up and
evolution of gas
and dust structures in the nuclear region of active galaxies, as well as the
origin of the dust content is still missing. 
The main reasons are the complexity of the physical processes, which are thought
to happen in these regions and their scale lengths, which make them invisible
for direct observations in most wavebands. 
However, a scenario of co-evolution of nuclear starbursts and tori, 
based on the models of \citet{Vollmer_04}, \citet{Beckert_04} and
observations of \citet{Davies_07} was suggested by \citet{Vollmer_08}. An external mass accretion 
rate causes three phases: (i) massive gas infall leads to a turbulent, stellar wind-driven $Q\approx1$ disc, 
where $Q$ is the Toomre stability parameter,  
(ii) subsequent supernova type\,II explosions remove the intercloud medium and a geometrically thick 
collisional clumpy disc remains. In the final phase (iii) with a low mass accretion rate, the torus gets geometrically 
thin and transparent.
Hydrodynamical simulations of the effects of stellar feedback from a young nuclear star cluster as
ubiquitously found in Seyfert galaxies \citep{Davies_07} was investigated in an
exploratory study by \citet{Schartmann_09}. Turbulent mass input in combination with supernova feedback and an 
optically thin cooling curve yield a two-component structure, made-up of a geometrically thick filamentary torus
component on tens of parsec scale and a geometrically thin, but turbulent disc structure. Subsequent continuum 
dust radiative transfer modelling results in a good comparison with available observational data sets. 
Given the success of these
simulations, this paper is a follow-up, concentrating more on the
long-term behaviour of this scenario and justifying our assumptions with the
help of further data comparison. 
This is enabled by treating the innermost part with a simplified
effective disc model and by concentrating on the nearby and well-studied Seyfert\,2 galaxy NGC\,1068.

After shortly reviewing the underlying global physical model
and its numerical realisation, the resulting state of the gas is discussed
briefly. The gas inflow of these simulations 
is used in a subsequent effective 
treatment of the resulting gas surface density of the inner disc
structure for a long-term evolution study. 
This enables us to link the large and the small scale structure
in these objects.
The comparison of feeding
parameters as well as structural parameters with observations and previous
modelling is done to assess that this model is a viable 
option and able to explain the mass assembly in the disc, the disc 
properties as well as the feeding of the central supermassive black hole.
Sect.~\ref{sec:discussion} provides a
critical discussion, before we summarise our
results in Sect.~\ref{sec:conclusions}.

\section{Large scale hydrodynamical torus model}

\subsection{Model and numerical realisation}
\label{sec:mod_numres}

In recent years, there has been growing evidence for a direct link between galactic nuclear activity and 
star formation on the parsec scale \citep{Davies_07}. 
Massive nuclear star clusters ($\approx 10^{8}\,M_{\odot}$), as ubiquitously found in Seyfert galaxies
significantly impact their environment during many stages of the evolution of
their stellar population. Analytical arguments based on energy considerations \citep{Davies_07} as well as 
numerical simulations \citep{Schartmann_09} reveal that during
the first phase of the evolution of the most massive stars towards core
collapse supernova, most of the gas will be expelled from the region of the stellar
cluster. A less violent phase follows, where the mass input is dominated by low
velocity winds during the {\it Asymptotic Giant Branch} (AGB) phase, after approximately
50 Myrs. At roughly this time, \citet{Davies_07} observationally find a switch-on process of
the active nucleus, interpreted as the beginning of accretion onto the black
hole of these low-velocity stellar ejecta. 
This is the time when our simulations start. The initial condition
comprises of a very low density, equilibrium distribution of matter for the 
given gravitational potential, which is washed out soon and is insignificant for
the further evolution. 
The injection of mass into the model space in our
simulations is mimicking the process of slowly expanding planetary nebulae
shells: a mass of $0.5\,M_{\odot}$ is evenly distributed within a
radius of 1~pc together with the equivalent of the kinetic energy of an
expanding shell of 30~km/s in form of thermal energy and with a bulk velocity
comprised of the (constant) random velocity and the rotational velocity 
(described by a power law) of the underlying stellar
distribution.
The nuclear star cluster in NGC\,1068 has a 
velocity dispersion of roughly 100\,km/s \citep{Davies_07}. 
Therefore, random motions are 
an important stabilising mechanism and the bulk rotation must be 
sub-Keplerian. For the parameters of our standard model,
the rotation velocity drops from approximately
70 km/s to about 30 km/s in the tens of parsec distance range, 
which corresponds well to the 
values derived from the SINFONI observations, which scatter around roughly
45 km/s \citep{Davies_07}.
The mass input rate is chosen according to the formalism described in \citet{Jungwiert_01}.  
Gas cooling is calculated with the help of an effective cooling
curve with solar metallicity, prepared with the help of the Cloudy code 
(\citealp{Plewa_95}, see also \citealp{Schartmann_09}, Fig.\,1). The hydrodynamical evolution is then followed within the
fixed gravitational potential of the nuclear stellar cluster, modelled with the help of
a Plummer distribution of stars and the Newtonian potential of a central
supermassive black hole with an influence radius of
$r_{\mathrm{BH}}=\frac{G\,M_{\mathrm{BH}}}{\sigma_*^2}\approx 3.4\,$pc.   
We use outflow boundary conditions in radial as well as polar direction, not allowing 
for inflow. Periodic boundary conditions are applied in the azimuthal direction, where
we only model a $90\degr$ fraction of the whole model space.

The hydrodynamics themselves are treated with the help of the {\sc Pluto} code
\citep{Mignone_07}. Being a fully MPI\footnote{Message Passing Interface}-parallelized 
high resolution shock capturing scheme with a
large variety of Riemann-solvers, we consider it perfectly suited to treat our
physical setup. For the calculations shown in this paper, we used the two-shock Riemann solver together with
a linear reconstruction method and directional splitting on a three-dimensional spherical coordinate system.

A more detailed description of the underlying physical model can be 
found in \citet{Schartmann_09}.
However, mind that due to large uncertainties in the delay times and the rates
of supernova type~Ia explosions (see discussion in \citealp{Schartmann_09}), in
this paper we solely concentrate on the effect of turbulent mass input of the nuclear 
star cluster.
The basic physical parameters of our simulations are summarised in Table~\ref{tab:param}. 
They are chosen to represent the nearby -- and therefore well studied -- Seyfert~2 galaxy NGC~1068.

\begin{table}
\begin{center}
\caption[Parameters of our standard 3D hydrodynamical NGC~1068 model]{Parameters of our standard hydro model.}
 \label{tab:param}
\begin{tabular}{lll}
\hline
Parameter & Value & Reference \\
\hline
$M_{\mathrm{BH}}$ & $8\,\cdot 10^{6}\,M_{\sun}$ & L03\\
$M_{*}$ & $2.2\,\cdot 10^{8}\,M_{\sun}$ & D07\\
$M_{\mathrm{gas}}^{\mathrm{ini}}$ & $1.0\,\cdot 10^{2}\,M_{\sun}$ & \\ 
$R_{\mathrm{c}}$ & 25\,pc & G03\\
$R_{\mathrm{T}}$ & 5\,pc & \\
$R_{\mathrm{in}}$ & 0.2\,pc & \\
$R_{\mathrm{out}}$ & 50\,pc & \\
$\sigma_{*}$ & 100\,km/s & D07\\
$\beta$ & 0.5 & \\
$T_{\mathrm{ini}}$ & $2.0\,\cdot 10^{6}\,$K  & \\
$\dot{M}_{\mathrm{n}}$ & $9.1\,\cdot 10^{-10}\,M_{\sun}/(yr\,M_{\sun})$ & J01\\
$M_{\mathrm{PN}}$ & $0.5\,M_{\sun}$ & \\
$\Gamma$ & $5/3$ & \\
\hline
\end{tabular}
\end{center}

\medskip
 Mass of the black hole 
 ($M_{\mathrm{BH}}$), normalisation constant of the stellar potential ($M_{*}$), 
 initial gas mass ($M_{\mathrm{gas}}^{\mathrm{ini}}$), cluster core 
 radius ($R_{\mathrm{c}}$), torus radius ($R_{\mathrm{T}}$), inner radius ($R_{\mathrm{in}}$), 
 outer radius ($R_{\mathrm{out}}$),
 stellar velocity dispersion ($\sigma_{*}$), exponent of the angular momentum
 distribution of the stars ($\beta$),
 initial gas temperature ($T_{\mathrm{ini}}$), normalised mass injection rate ($\dot{M}_{\mathrm{n}}$), 
 mass of a single injection ($M_{\mathrm{PN}}$) 
 and adiabatic exponent ($\Gamma$). The references are: L03 \citep{Lodato_03},
 D07 \citep{Davies_07}, G03 \citep{Gallimore_03} and J01 \citet{Jungwiert_01}. 
 For a detailed description of the model,
 we refer to \citet{Schartmann_09}.
\end{table}

\subsection{Resulting density and temperature distribution}

\begin{figure*}
\begin{center}
\includegraphics[width=0.7\linewidth]{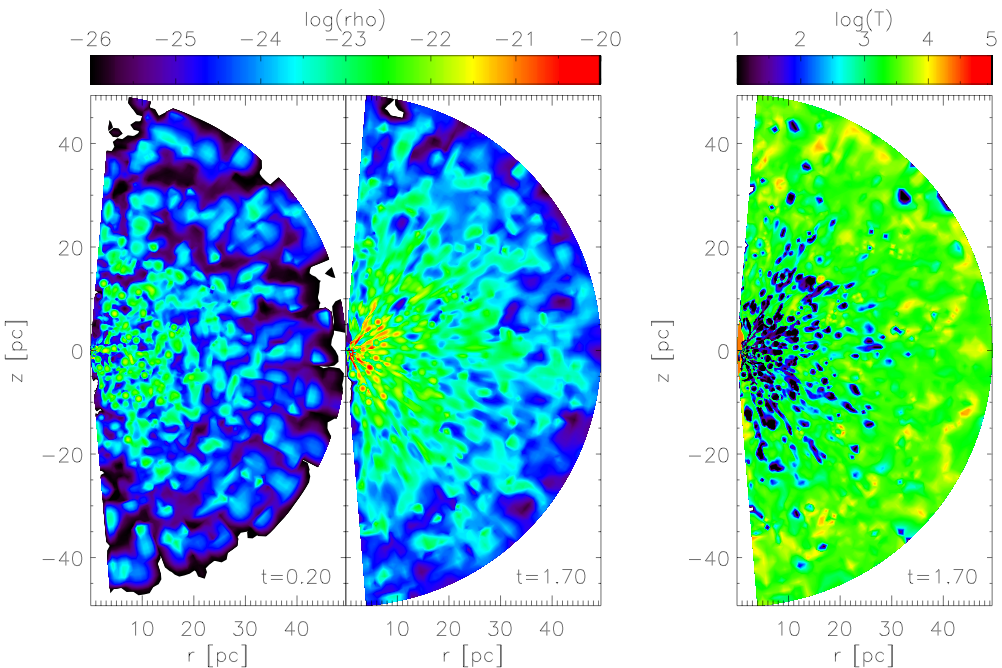}
\caption{Snapshots of the density distribution in a meridional
  plane after (a) 0.2 orbits (corresponding to $7 \cdot 10^4$\,yr) and (b) 1.7 orbits
  (approximately $6 \cdot 10^5$\,yr). The panel on the right hand side (c) shows the
  temperature distribution in the same meridional plane after 1.7 orbits.}
% figure produced with 
% ex-srv4:~/data/routines/IDL/PLUTO_IDL/den_temp_evol.pro
% in:
% ex-srv4:~/data/LRZ_copy/2009/global_tor_highres_000/
\label{fig:den_temp_evol}
\end{center}
\end{figure*}

The resulting temporal evolution of the density distribution within a
meridional plane in a three-dimensional high-resolution simulation 
($n_{\mathrm{r}}=250$, $n_{\theta}=133$, $n_{\phi}=70$) with the parameters 
given in Table~\ref{tab:param} as well as 
the same cut through the temperature distribution is shown in Fig.~\ref{fig:den_temp_evol}.
In the first time steps of the simulation, the density distribution is
dominated by the original, injected mass blobs. Due to their large number
density within the domain and their high randomly oriented velocities, 
collisions are very frequent.
In our pure hydrodynamical setup, these collisions are inelastic. Therefore,
the blobs dissipate a large fraction of their kinetic energy -- which is 
radiated away due to optically thin line and continuum cooling processes -- and many of them
merge to form larger complexes.
As discussed in Sect.~\ref{sec:mod_numres}, the stars (and therefore also the
clouds at the time of injection) possess sub-Keplerian rotation velocities at 
the tens of parsec distance from the centre. Therefore, the larger gas clouds 
no longer possess stable orbits, after 
losing a certain amount of kinetic energy (from the random component mainly),
and tend to stream radially inwards, where they find their new equilibrium 
at their angular momentum barrier, after suffering some angular momentum 
redistribution on their way towards the centre. 

\begin{figure}
\begin{center}
\includegraphics[width=0.9\linewidth]{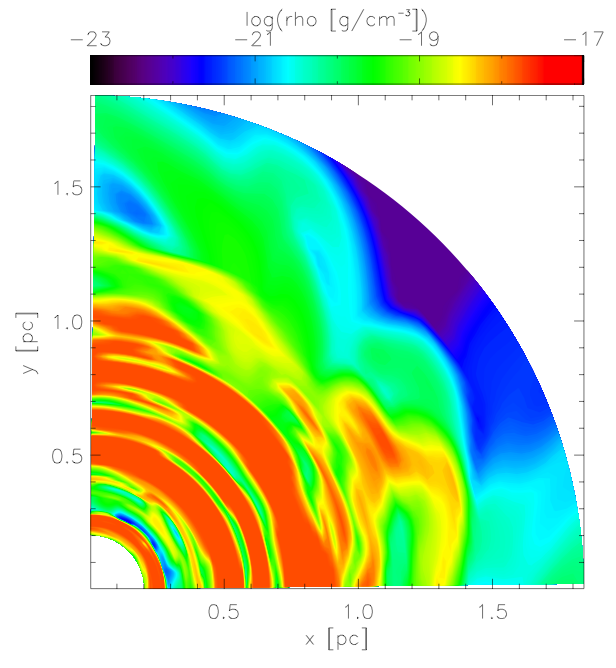}
\caption{Cut through the innermost part of the equatorial plane of our 3D
  hydrodynamical standard model after $6 \cdot 10^5$\,yr, 
  showing the nuclear disc component.}
% figure produced with 
% ex-srv4:~/data/routines/IDL/PLUTO_IDL/den_inner_disk_onesnap.pro
% in:
% ex-srv4:~/data/LRZ_copy/2009/global_tor_highres_000/
\label{fig:den_inner_disc}
\end{center}
\end{figure}

In the course of the simulation, a two-component structure builds up: a
large-scale clumpy or filamentary component (mainly visible in the plots of
Fig.~\ref{fig:den_temp_evol}) and a geometrically thin, but very dense disc on parsec scale 
(not visible in the plots)\footnote{To avoid confusion: The two-component
structure is what we called the torus in Sect.\ref{sec:introduction}. 
When we talk about {\it discs} in the following, we always refer to this dense 
nuclear disc component on parsec scale. The accretion disc in the immediate vicinity 
of the black hole (ranging from the marginally stable orbit up to a few thousand 
Schwarzschild radii) will be called {\it hot inner accretion disc} from now on.}. 
Clumps and filaments from the outer component fall
onto the thin disc and cause some amount of turbulence, triggering angular
momentum transport outwards and accretion\footnote{Please mind that the term
{\it accretion} not necessarily refers to accretion onto the central black
hole, but it also means simply mass transfer through the nuclear disc in our
terminology.} inwards. 
The corresponding temperature distribution is depicted in Fig.~\ref{fig:den_temp_evol}(b) 
and shows complementary behaviour with respect to the density distribution: 
the densest blobs are the coldest, whereas the tenuous gas in-between is the hottest. 
This is due to the optically thin gas cooling, which scales quadratically 
with the density distribution.
Fig.~\ref{fig:den_inner_disc} shows the innermost part of this
simulation in a cut through the equatorial plane. The main disc component in our
simulation spans a radial
range between roughly 0.5\,pc and 1.1\,pc. 
This is in approximate agreement with a disc found with the help of water maser
observations with an extent between 0.65\,pc to 1.1\,pc \citep{Greenhill_97}. 
This shows that the angular momentum distribution of the gas flowing towards the
centre seems to be reasonable.
Note however that 
the outer large scale torus part is in equilibrium already, but 
mass still assembles in the disc component of our simulations.   
Although filling most of the volume of the model space, the obscuration
properties are mostly given by the inner dense disc component. The outer
fluffy torus part alone can not account for the Seyfert 1/2 dichotomy. 

However, it is still a matter of debate what the actual driver for angular momentum 
redistribution in such discs is. Due to this missing mechanism and the lack of 
feedback due to the formation of stars in our simulations, more and more gas from 
the large-scale torus component assembles in the disc, which prevents us from 
treating the whole Seyfert activity cycle -- which is expected to be of the order of
$10^7$ to $10^8$ years  -- in one simulation. 
Apart from this, these kind of computations are very time consuming. 

Therefore, we use our 3D hydrodynamical
simulations to describe the large-scale structure and attach an axisymmetric
viscous disc model at small scales. In this paper, we are especially interested in
comparing the resulting neutral hydrogen column densities, disc sizes and
accretion rates through our inner boundary to various observations.

\section{The effective nuclear disc model}
\label{sec:discmodel}

\subsection{Overall model and numerical realisation}

The idea behind these simulations is to use the gas supplied by our large
scale three-dimensional {\sc Pluto} torus simulations, which is driven inwards due to dissipation of
turbulent cloud motions in collisions. 
The resulting dense clumps cool on small time scale, acting as a stabilising
mechanism for the clouds and preventing them from smearing out into a
continuous medium, which happens as soon as cooling is turned off. 
Additionally, some amount of angular momentum is transported with the help of these
collisions. This inward motion stalls at the respective radius where the
angular momentum barrier is reached for the corresponding gas clump and
leads to the formation of a nuclear gas disc. Being
restricted by the physics we are able to take into account in the large scale
simulations, this disc continuously grows in mass in our simulations, with
only a very small amount being transported inwards. Simplified 
effective disc simulations will help us in the following to test, whether a stationary state
can be reached and how the physical parameters of these discs compare to
observations. In this 
effective disc model we calculate the
viscous evolution of such discs including star formation and taking mass input
from our large-scale simulations into account. Angular momentum is mediated 
by a turbulent viscosity (e.~g.~ generated by magneto-rotational instability 
or a self-gravitating disc) and star formation happens in the
densest regions of the disc.   

Following \citet{Pringle_81} and \citet{Lin_87}, the viscous evolution of such
a disc can be described by the following differential equation:

\begin{eqnarray}
\label{equ:dgl}
\frac{\partial}{\partial t} \Sigma(t,R) + \frac{1}{R} \, \frac{\partial}{\partial
  R} \, \left[\frac{\frac{\partial}{\partial R} \left(
      \nu_{\alpha} \, \Sigma(t,R) \, R^3 \, \frac{d \Omega(R)}{d R}
    \right)}{\frac{d}{d R} \left( R^2 \, \Omega \right)} \right]
= S 
\end{eqnarray}

This equation is derived from mass and angular momentum conservation, 
coupled via the radial velocity (for details see \citealp{Pringle_81}).  
$\Sigma(t,R)$ is the surface density depending on time $t$ and radius $R$
within the disc. 
The rotation is Keplerian yielding the following rotation
frequency:

\begin{eqnarray}
  \Omega = \sqrt{G \, M_{\mathrm{BH}}} \, R^{-1.5}
\end{eqnarray}
with the mass of the nuclear supermassive black hole $M_{\mathrm{BH}}$
and $G$ being the gravitational constant. 
The stellar mass component is neglected here, as we are well inside the influence radius
of the black hole ($\approx 3.4\,$pc for NGC\,1068).
We assume that the viscosity is due to some turbulent process and describe it
with the help of a turbulent $\alpha$ viscosity \citep{Shakura_73}:

\begin{eqnarray}
  \nu_{\alpha} = \alpha_{\nu} \, c_{\mathrm{s}} \, H(R),
\end{eqnarray}

where $\alpha_{\nu}$ is a dimensionless parameter determining the strength of
the angular momentum transport ($\alpha_{\nu} \in [0,1]$), $c_{\mathrm{s}}$ is
the speed of sound and $H(R)$ is the scale height of the disc

\begin{eqnarray}
\label{equ:scaleheight}
H(R) = \delta\,R,
\end{eqnarray} 

with the scaling parameter $\delta$.
We added the extra terms abbreviated by $S$. They include a source of mass
due to the input from large scale ($\dot{\Sigma}_{\mathrm{input}}(t,R)$, 
see Sect.~\ref{sec:massin}) and a sink due to
star formation ($\dot{\Sigma}_{\mathrm{SF}}(t,R)$, see Sect.~\ref{sec:sftreat}): 

\begin{eqnarray}
S = \dot{\Sigma}_{\mathrm{input}}(t,R) - \dot{\Sigma}_{\mathrm{SF}}(t,R)
\end{eqnarray}

Equation \ref{equ:dgl} is solved with the help of 
{\sc Matlab}'s\footnote{http://www.mathworks.com/products/matlab/} pdepe solver. 
A summary of the parameters of our effective disc standard model
in addition to the ones used for the 3D {\sc Pluto} calculations are 
summarised in Table~\ref{tab:param_disc}.

\begin{table}
\begin{center}
\caption[Parameters of our standard efective disc model]{Parameters of our standard
  effective disc model.}
 \label{tab:param_disc}
\begin{tabular}{ll}
\hline
Parameter & Value \\
\hline
$R_{\mathrm{in}}$ & $0.1\,$pc \\
$R_{\mathrm{out}}$ & $100.0\,$pc \\
$\delta$ & 0.2 \\
$T$ & 400\,K \\
$\alpha_{\nu}$ & 0.05 \\
$t_{\mathrm{cluster}}^{\mathrm{start}}$ & $50\,$Myr \\
$t_{\mathrm{cluster}}^{\mathrm{end}}$ & $300\,$Myr \\
$n_{\mathrm{r}}$ & 5000 \\
\hline
\end{tabular}
\end{center}
\medskip
 Inner radius of the domain ($R_{\mathrm{in}}$), outer radius of the domain
 ($R_{\mathrm{out}}$), thickness of the disc ($\delta$ = disc scale height / radius of the disc),
 gas temperature ($T$), $\alpha$ viscosity parameter ($\alpha_{\nu}$), 
 age of the nuclear star cluster at the beginning of the
 simulations ($t_{\mathrm{cluster}}^{\mathrm{start}}$) and at the end
 ($t_{\mathrm{cluster}}^{\mathrm{end}}$) and resolution of the simulations ($n_{\mathrm{r}}$). 
\end{table}

\subsection{Mass input from our Turbulent Torus Simulation}
\label{sec:massin}

\begin{figure}
\begin{center}
\includegraphics[width=0.98\linewidth]{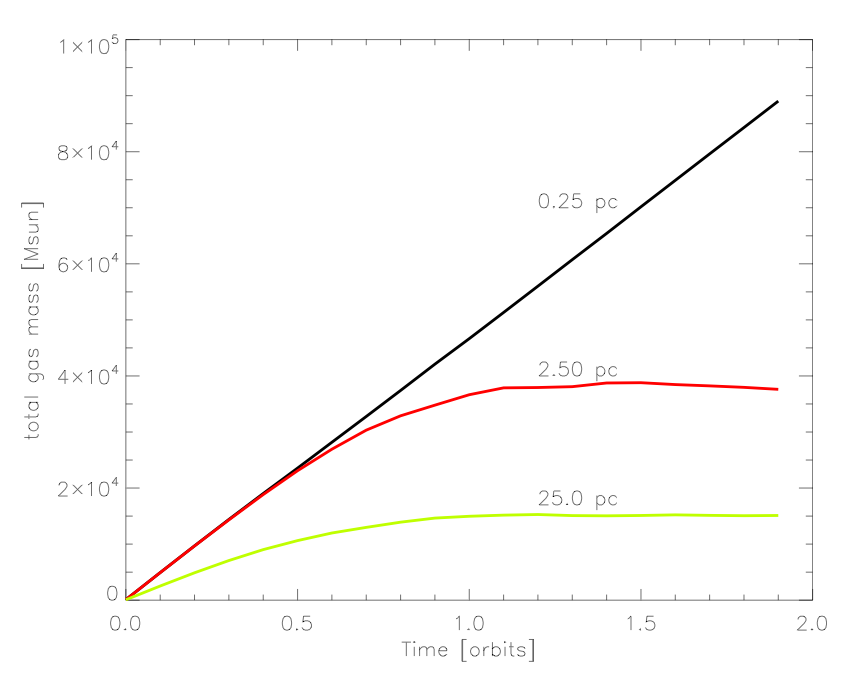}
\caption{Total gas mass within the domain of our 3D PLUTO simulations outside a sphere of radius 0.25\,pc,
  2.5\,pc and 25\,pc.}
% figure produced with 
% ex-srv4:~/data/routines/IDL/PLUTO_IDL/fig_input_radii.pro
% in ex-srv4:~/data/LRZ_copy/2009/global_tor_highres_000/
\label{fig:Minput_radii}
\end{center}
\end{figure}

%We use the simulation tor\_highres\_log\_copy in 
%h0075ad@a01:/ptmp2/h0075/h0075ad/2009/tor\_highres\_log\_copy
%which calculates the outer part of our turbulent torus model. 
Already after the comparatively short evolution time of our three-dimensional hydrodynamical
simulations, the outer
part of the simulation has reached a stable equilibrium, concerning
the total mass balance. This can be
seen in Fig.~\ref{fig:Minput_radii}, where the total gas mass outside a given
radius (0.25\,pc, 2.5\,pc and 25.0\,pc) is plotted against the time.
Whereas the smallest radius shown lies within the dense disc structure, we are
well outside the disc further out than 2.5\,pc in radius. 
Therefore, we use all gas, which is advected through a sphere of radius $2.5\,$pc
for our effective disc simulations. 
This amounts to gas mass transfer of $0.14\,M_{\odot}$/yr at a cluster age of $50$\,Myr. 
The temporal evolution of the accumulated mass of this gas is shown as the red
dashed line in Fig.~\ref{fig:sim1_mass_alpha} and Fig.~\ref{fig:sim1_mass_temp}. The
change of slope is due to the decreasing mass input rate with time, according
to the analytical prescription of the mass loss processes of a stellar
population as described by \citet{Jungwiert_01}. 
The radial location of input is chosen such that the angular momentum of the 
gas coming from the large scale hydrodynamical simulations equals the angular
momentum of a Keplerian disc at the respective radius:

\begin{eqnarray}
r_{\mathrm{input}} = \frac{j_{\mathrm{gas,input}}^2}{G\,\left(M_{\mathrm{BH}}
    + M_{\mathrm{*}}(R)\right)},   
\end{eqnarray}
where $j_{\mathrm{gas,input}}$ is the specific angular momentum of the gas.

\begin{figure}
\begin{center}
\includegraphics[width=0.98\linewidth]{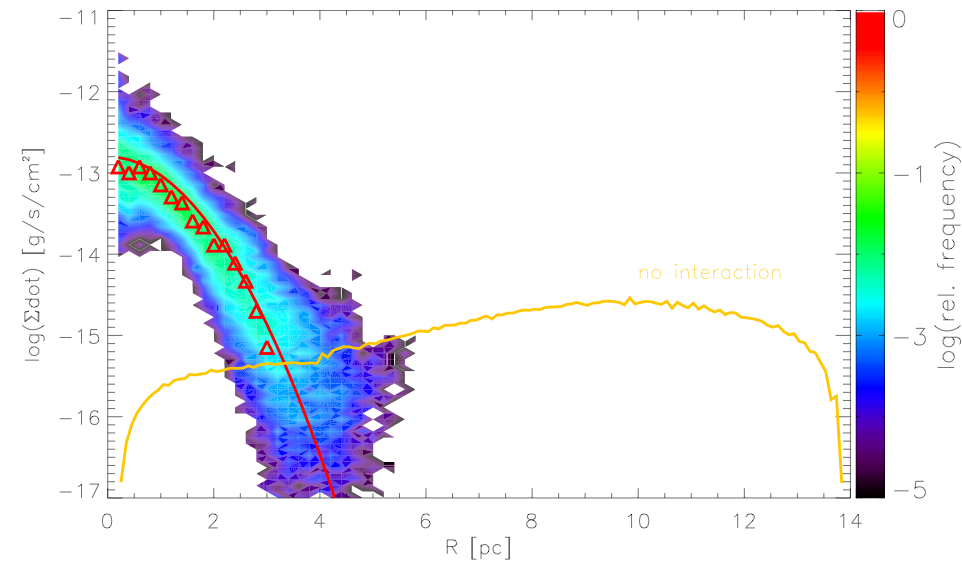}
\caption{Mass input into our effective disc model,
  shown as the growth of the gas surface density 
  at different radii, 
  flowing in through a sphere with radius of $2.5\,$pc from our 3D {\sc Pluto} turbulent torus simulation. 
  The input radii are set according to an angular momentum
  distribution of a Keplerian disc. Shown is the logarithm of a histogram for 100
  timesteps with a resolution of approximately 3400\,yr between an evolution
  time of 0.7\,Myr and 1.0\,Myr, where the 3D hydro simulation is already in an
  equilibrium state. The red triangles denote the maxima of each radial bin, the red
  parabola is the parametrisation used as source term in the effective disc simulations. The yellow
  graph denotes the growth of the gas surface density, if we directly take the angular momentum
  of the mass blobs, ejected from the stellar cluster and do not evolve them in a hydrodynamical 
  simulation.}
% figure produced with:
% ex-srv4:~/data/routines/IDL/PLUTO_IDL/disk_matlab_input_new_cool.pro
% in ex-srv4:~/data/LRZ_copy/2009/tor_highres_log_copy_cool/
\label{fig:mdot_input_histo_025}
\end{center}
\end{figure}

This assumes that mass inflow does not interact with disc material before settling into 
the disc plane (e.~g.~gas inflow happens from all directions and the disc is geometrically thin).
Fig.\,\ref{fig:mdot_input_histo_025} 
shows the resulting histogram of the mass flow through a sphere of radius
2.5\,pc from the {\sc Pluto} simulation. The histogram is 
calculated from a slightly higher resolved {\sc Pluto} run with a resolution of
$n_{\mathrm{r}}=200$, $n_{\mathrm{\theta}}=197$ and $n_{\mathrm{\phi}}=103$, which was restricted
to a radial range between 2.5\,pc and 50\,pc. The time
resolution of the gas flow derivation is approximately 3400\,yr. For the
histogram, a total of 100 snapshots have been used between an evolution
time of 0.7\,Myr and 1.0\,Myr. 
Colour coded is the relative frequency of a given combination of radius and surface density growth rate
within the 100 snapshots. 
For each radial bin of this histogram, the red triangle denotes the maximum
of the respective radial bin (the most frequent combination of this radius and mass
input rate of the snapshots). We then fit these triangles with a parabola and renormalise to the total mass input rate. This parametrised mass
input rate is given by 

\begin{eqnarray}
\label{equ:mdotin}
\dot{\Sigma}_{\mathrm{input}}(t,R) = 10^{-12.80-0.23\,R[pc]^2}
\end{eqnarray}

where the radius in the disc $R$ is given in units of parsec as indicated.
This function, which is also shown as the red parabola in Fig.~\ref{fig:mdot_input_histo_025}, 
is used in the simulations of this paper. 
Mind that a small portion of the gas at $2.5\,$pc possesses more angular
momentum than it would have in a Keplerian disc at the same radius.

As already mentioned before, the stars in the nuclear cluster rotate with sub-Keplerian 
velocities. 
If we would use the mass injections of the stars within the nuclear star cluster directly
and feed them into a Keplerian rotating disc, without hydrodynamically calculating the interaction of the 
blobs, then the yellow graph would result. 
However, during the 3D PLUTO simulations, turbulent motions within the large scale torus component transport angular momentum
outwards and lead to accretion of matter and therefore to a further concentration of the 
disc towards the centre, evident in the different shapes of the red and yellow graph.

\subsection{Treatment of star formation}
\label{sec:sftreat}

Cold nuclear discs of this kind are known to be able to form stars 
\citep{Paumard_06,Nayakshin_06a,Alig_09}.
In our simulations, we convert gas into stars, whereever the Toomre stability
parameter 

\begin{eqnarray}
Q = \frac{c_{\mathrm{s}}\,\Omega}{\pi\,\Sigma\,G}
\end{eqnarray}

is smaller than unity, where $\Omega$ is the Keplerian orbital frequency,
$c_{\mathrm{s}}$ is the local speed of sound, $\Sigma$ is the gas surface
density and $G$ is the gravitational constant. The amount of gas removed from
the computational domain is chosen according to the Kennicutt-Schmidt star
formation law \citep{Kennicutt_98}: 
\begin{eqnarray}
\dot{\Sigma}_{\mathrm{SF}} = 2.5\cdot 10^{-4} \, \left[ \frac{\Sigma}{\frac{M_{\odot}}{\mathrm{pc}^2}}
\right]^{1.4} \, \frac{M_{\odot}}{\mathrm{yr}\,\mathrm{kpc}^2}
\end{eqnarray}

$\Sigma$ is as usual the gas surface density within the disc. 
Gas which forms stars is simply removed from the simulation and will no longer
participate in the viscous evolution of the disc.

\subsection{Results of the effective disc models and comparison with observations}
\label{sec:res_effdisc}

%\subsubsection{Varying the strength of angular momentum transfer}
%\label{sec:alpha_study}

\begin{figure}
\begin{center}
\includegraphics[width=0.98\linewidth]{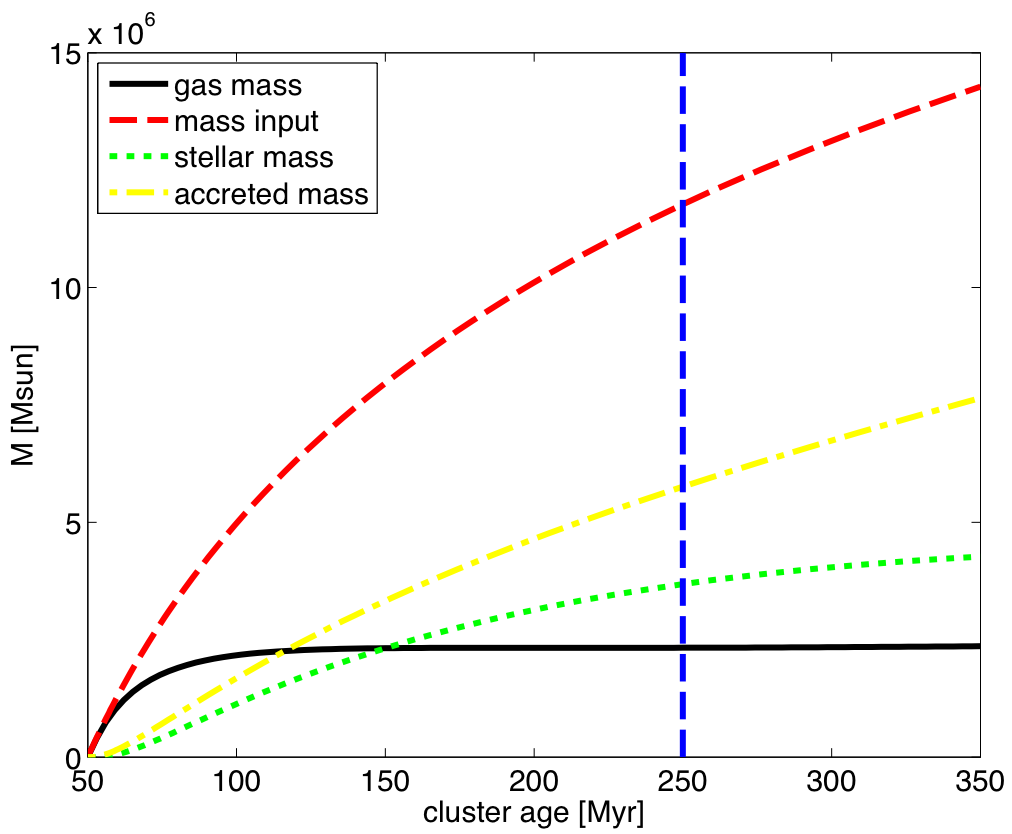}
\caption{Comparison of mass contributions of the various components of our
  model for an $\alpha$ viscosity parameter of 0.05 and a gas temperature of 400\,K.
  The blue dashed line denotes the estimated current age of the nuclear star cluster in NGC\,1068.}
% figure produced with:
% plot_study.m
% in mschartm@ex-sol:~/calculations/disk_evolution/paper_const_scaleheight_revised/alpha_study_kepler/
\label{fig:sim1_mass_alpha}
\end{center}
\end{figure}

In the course of the simulation, several physical processes are competing with
one another. Starting with a basically empty model space in all our
simulations, the knowledge of the 
initial condition is lost on very short timescale.  
The mass inflow from evolving stars in our large scale simulation
first builds up the rotating disc, which is then constantly replenished over
time, but with a decreasing mass input rate. The turbulent viscosity tries to
move angular momentum outward, 
leading to accretion of gas through the disc towards the inner edge and
beyond. The residual of the newly introduced gas remains in the disc and at some
point, when the Toomre stability criterion is not fulfilled anymore, gravitational
collapse can occur and stars will form, contributing to a stellar disc, which
then coexists with the gas structure. 
A comparison between the various contributions for our standard model
(for the parameters given in Table\,\ref{tab:param} and \ref{tab:param_disc}) is
shown in Fig.\,\ref{fig:sim1_mass_alpha} as a function of time.   
The vertical dashed blue line marks the estimated current age of the nuclear star cluster in
NGC\,1068 of 250\,Myr \citep{Davies_07}. The integrated
gas mass introduced into 
the disc simulation is given as red dashed line. Most of this mass then gets accreted
through viscous processes, as shown with the yellow dash-dotted line. 
The gas residing in the disc is given as black solid line. It rises steeply during
the first few Million years. By then, gas transport through the inner boundary 
and transformation of gas into stars
get significant, leading to a turn over of the disc
mass curve. 
In the further evolution, the disc mass stays approximately constant, whereas the slopes 
of the total accreted gas mass as well as the total mass in stars flatten due to the
decreasing mass inflow rate. 

\begin{figure}
\begin{center}
\includegraphics[width=0.98\linewidth]{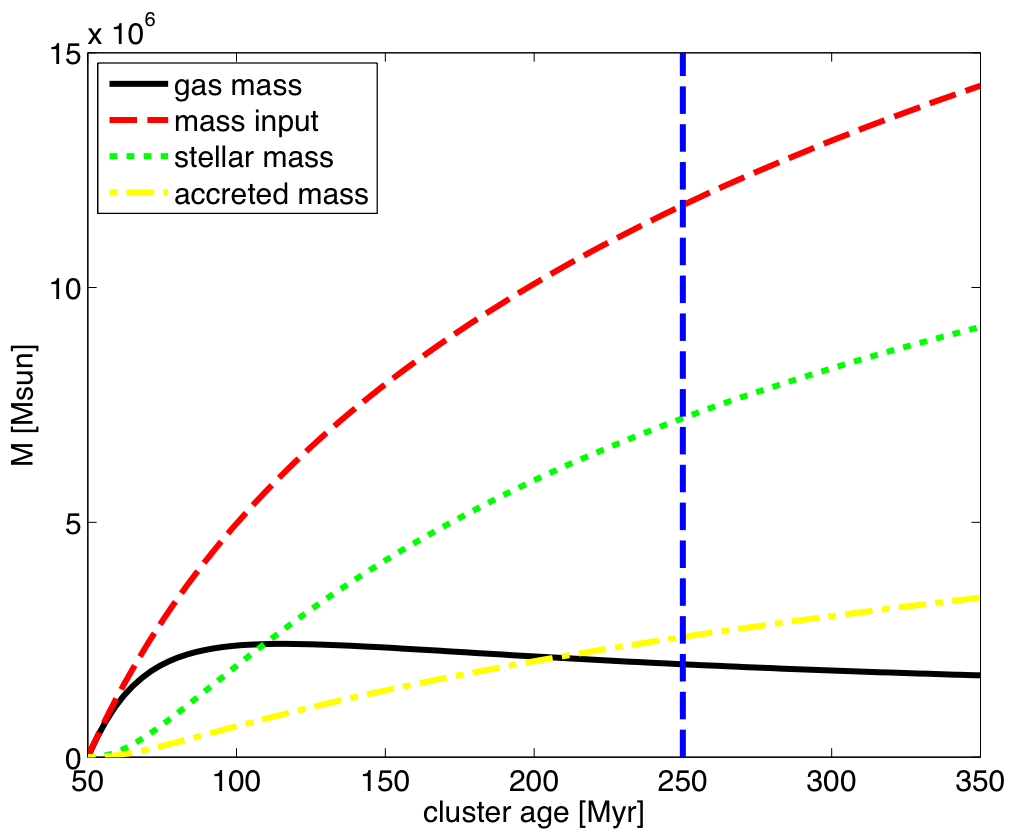}
\caption{Comparison of mass contributions of the various components of our
  model for an $\alpha$ viscosity parameter of 0.05 and a temperature of 50\,K.}
% figure produced with:
% plot_study.m
% in mschartm@ex-sol:~/calculations/disk_evolution/paper_const_scaleheight_revised/temp_kepler/
\label{fig:sim1_mass_temp}
\end{center}
\end{figure}

The same comparison of contributions to the total integrated mass budget, but for the
case of a lower gas temperature of only 50\,K is shown in 
Fig.~\ref{fig:sim1_mass_temp}. 
When decreasing the gas temperature, the Toomre criterion indicates
instability already at lower gas densities. Therefore, star formation now dominates
over the gas mass residing in the disc after roughly 110\,Myr. 
It contributes the largest fraction of the inserted gas mass. 
The disc mass evolution now shows a visible decrease at later times. 
The dependence of star formation on the temperature of the gas in
the disc can be seen directly in the comparison between
Fig.~\ref{fig:sim1_mass_alpha} and Fig.~\ref{fig:sim1_mass_temp}, which shows
a difference of roughly a factor of two between the stellar mass in the 400\,K and the 50\,K
simulation. In reality, we expect the disc to consist of a multiphase medium, with regions
of cold, warm and even hot gas due 
to gas cooling and a variety of heating processes within the disc, so
that the star formation rate might be intermediate between the 
examples depicted here \citep[e.~g.~][]{Schartmann_09}.  

In the following part of this section, we study how these processes compete with
one another, when changing the (generally unknown) $\alpha$ viscosity
parameter of the disc. In this study, it takes the values 0.05, 0.1 and 0.2.
Observationally, \citet{King_07} find for thin, fully ionised accretion discs
a typical range of values between 0.1 and 0.4. Numerical studies of 
magnetohydrodynamical turbulence tend to
derive up to an order of magnitude smaller values. 
For the case of circum nuclear discs, which are only partially ionised, angular momentum
transport is reduced and depends strongly on the ionisation fraction of the gas.

\begin{figure}
\begin{center}
\includegraphics[width=0.98\linewidth]{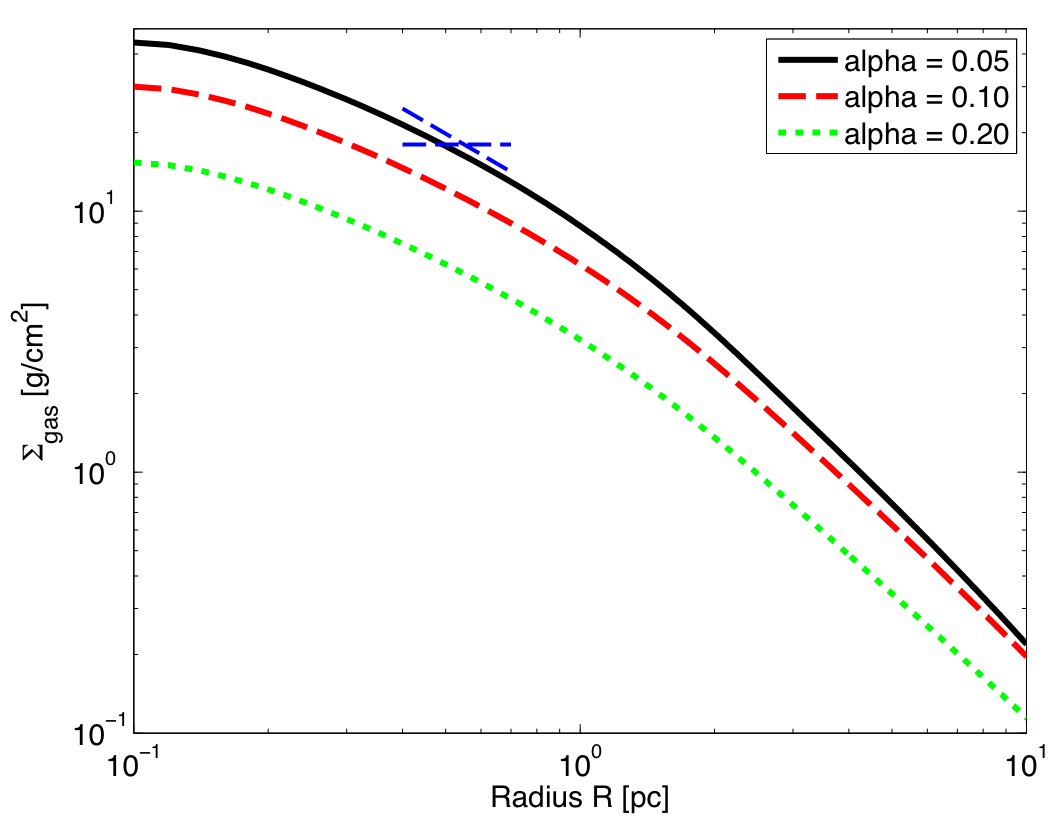}
\caption{Final gas density distribution at a cluster age of
  250\,Myr. Superposed as blue dashed lines are the two limiting curves for
  the range of possible surface density distributions as determined from MIDI
  observations \citep{Kishimoto_09}.
}
% figure produced with:
% plot_study.m
% in mschartm@ex-sol:~/calculations/disk_evolution/paper_const_scaleheight_revised/alpha_study_kepler/
\label{fig:gasfinal_alpha}
\end{center}
\end{figure}

\begin{figure}
\begin{center}
\includegraphics[width=0.98\linewidth]{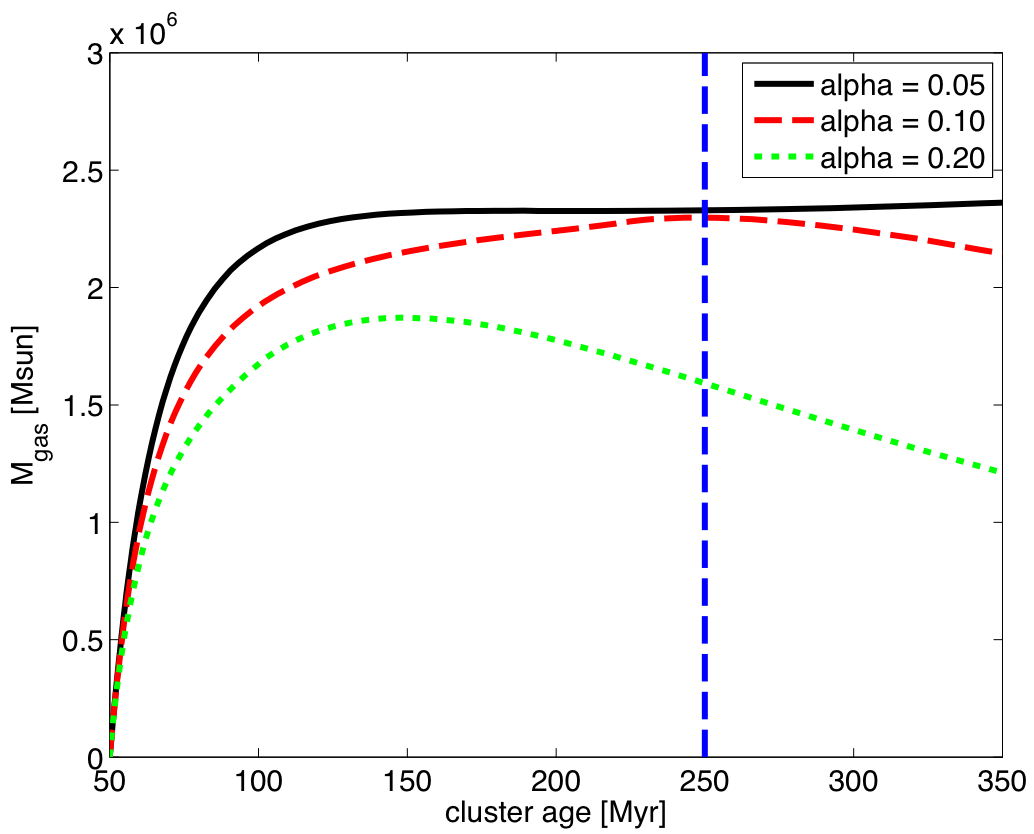}
\caption{Total gas mass of the disc.}
% figure produced with:
% plot_study.m
% in mschartm@ex-sol:~/calculations/disk_evolution/paper_const_scaleheight_revised/alpha_study_kepler/
\label{fig:mgas_tot_alpha}
\end{center}
\end{figure}

Fig.~\ref{fig:gasfinal_alpha} shows the gas surface density distribution for the
three $\alpha$ parameters: $\alpha_{\nu}=0.05$ as black solid line,
$\alpha_{\nu}=0.10$ as 
red dashed line and $\alpha_{\nu}=0.20$ as green dotted line at an age of the nuclear star
cluster of 250\,Myr. The qualitative difference between the
parameter study is as expected: the smaller the viscosity (scaling
proportional to 
$\alpha_{\nu}$), the more mass is able to assemble in the disc. This is also
reflected in the graph of the total gas mass in the disc
(Fig.~\ref{fig:mgas_tot_alpha}). 

In order to be able to compare the sizes of the discs with MIDI
observations, the following procedure is applied:
MIDI detects hot dust (see Sect.~\ref{sec:introduction}), which can only exist
beyond the sublimation radius, which was estimated to be roughly 
0.4\,pc for the case of NGC\,1068 \citep{Greenhill_96,Schartmann_05}. 
Therefore, we cut our discs at this radius,
assume a constant gas-to-dust 
ratio and derive the half width at half maximum (HWHM) of the remaining structure. 
This then yields a HWHM of the dust disc between approximately 0.84\,pc and
0.86\,pc for the three $\alpha$ parameters at
the current age of the nuclear star cluster in NGC\,1068. Remarkably, this is
very similar to the extent of the inner component as seen with MIDI. Here, a
half width at half maximum of approximately $0.7\,$pc is measured
\citep{Raban_09}. As already mentioned before, a warped disc as seen
in water maser observations of the same size was found by  
\citet{Greenhill_97}. It extents between 0.65\,pc and 1.1\,pc.
Water maser emission traces warm ($\approx 400\,$K), high-density
($n_{H_2}\approx 10^8-10^{10}cm^{-3}$) molecular
gas \citep{Elitzur_92}. 
However, one should note that it is still unclear
how the structure detected in hot dust emission with MIDI can be connected
to the very dense and clumpy disc assumed from
water maser emission. Nevertheless, the finding in different modes of observations
as well as in physically motivated simulations suggests a common origin.   

MIDI observations of a sample of nearby AGN suggest a common radial structure of AGN tori with a surface
brightness distribution proportional to $r^{-2}$, which corresponds roughly to a
(more or less model independent) surface density distribution proportional to
$r^{-1...0}$ \citep{Kishimoto_09}. Line segments with the two limiting
exponents (-1 and 0) are 
superposed as blue dashed line segments in the radial range of the
detection of a disc-like dust structure in NGC\,1068 by MIDI with a size of
roughly 0.7\,pc in radius \citep{Raban_09}. In this radial range, the shape of the 
surface density distribution of our discs is well consistent with these slopes inferred from
observations by \citet{Kishimoto_09}, but are rather at the upper end. 

The time evolution of the total gas mass within the disc
(Fig.~\ref{fig:mgas_tot_alpha}) shows a steep rise at the 
beginning of the simulation, until the gas disc is
build up and a maximum is reached. 
Then, it is followed by a shallow decay, due to the decreasing gas mass input
rate. Values are of the order of $2 \cdot 10^6\,M_{\odot}$ in the maximum. 
In good comparison to that, 
\citet{Kumar_99} found a disc mass of the order of $10^6\,M_{\odot}$ with the help
of a clumpy disc model, which accounts for the observed maser emission. 
Furthermore, a highly inclined thin, and clumpy ring or nuclear disc with
a molecular mass of $1.3\cdot 10^6\,M_{\odot}$ \citep{Montero_09} has been found in our own galactic 
centre in a distance ranging from approximately 2 to
5\,pc \citep{Genzel_85,Guesten_87}. 
This is particularly interesting, because the black hole mass in the galactic
centre of $4\cdot 10^6\,M_{\odot}$ \citep{Ghez_05,Schoedel_09} is very similar to
the one in NGC\,1068 and the mass of its circum nuclear disc is comparable to
the results of our infall model.    

\begin{figure}
\begin{center}
\includegraphics[width=0.98\linewidth]{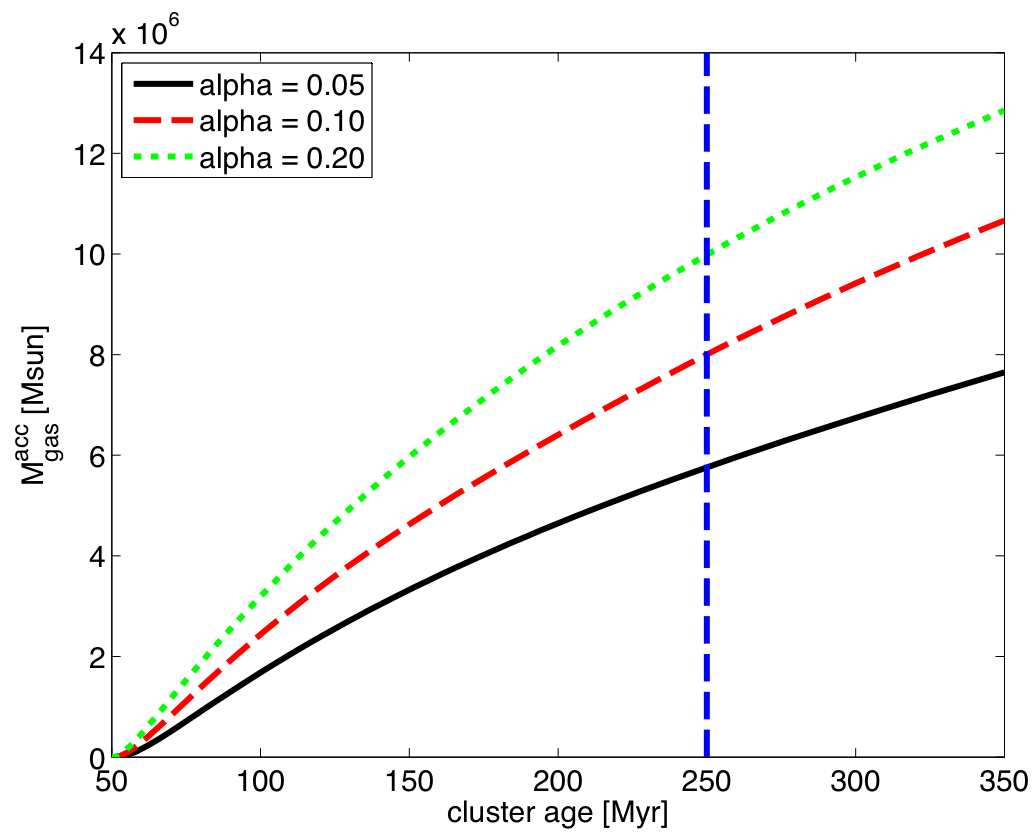}
\caption{Total accreted gas mass through the inner boundary of the domain.}
% figure produced with:
% plot_study.m
% in mschartm@ex-sol:~/calculations/disk_evolution/paper_const_scaleheight_revised/alpha_study_kepler/
\label{fig:macc_tot_alpha}
\end{center}
\end{figure}

\begin{figure}
\begin{center}
\includegraphics[width=0.98\linewidth]{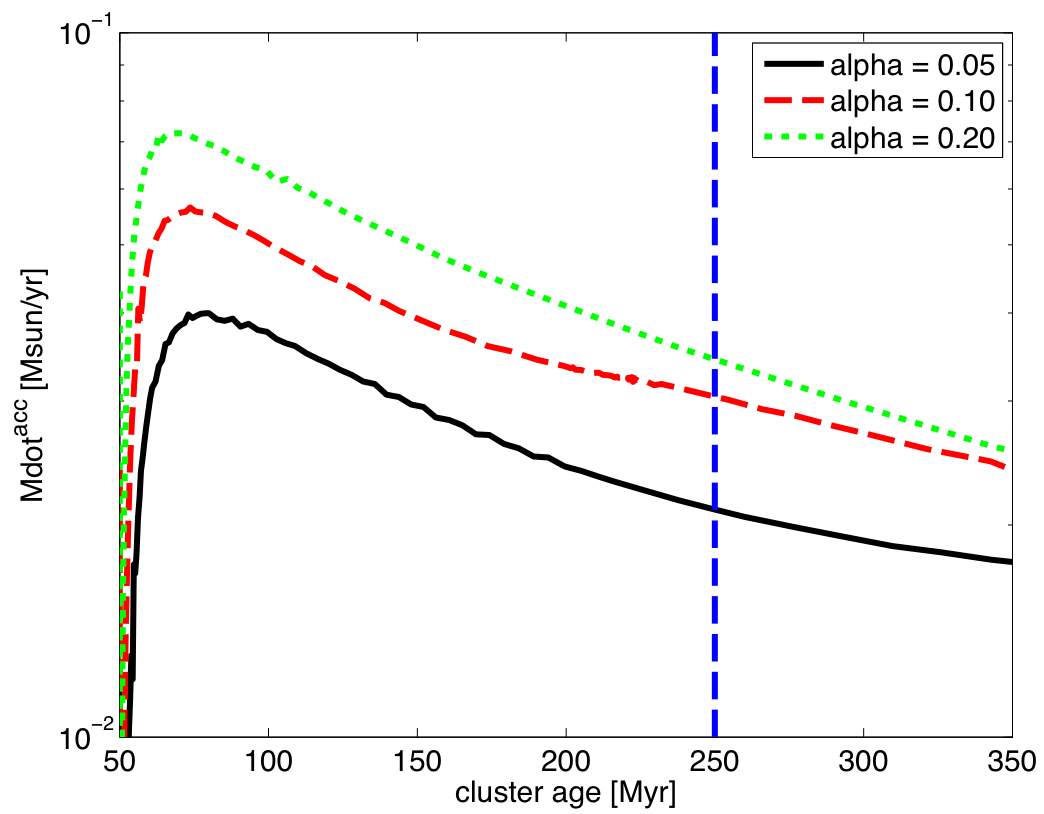}
\caption{Total accretion rate of gas through the inner boundary of the domain.}
% figure produced with:
% plot_study.m
% in mschartm@ex-sol:~/calculations/disk_evolution/paper_const_scaleheight_revised/alpha_study_kepler/
\label{fig:mdot_acc_alpha}
\end{center}
\end{figure}

Fig.~\ref{fig:macc_tot_alpha} shows the total accreted mass of the parameter
study. As expected, larger $\alpha_{\nu}$-values lead to a larger amount of
angular momentum transfer outwards and to a larger amount of gas accretion
through the inner boundary, which can also be seen in the current mass 
transfer rate through the inner boundary (Fig.~\ref{fig:mdot_acc_alpha}) in
solar masses per year as a function of time. 
After an evolution time of $250$\,Myr after the cluster
formation, a typical accretion rate of approximately $0.025\,M_{\odot}$/yr results. 
It is generally unknown, which fraction of this
mass transported through a sphere of radius 0.1\,pc really reaches the black
hole. Assuming that the efficiency is somewhere in between 10\,\% and 100\,\%,
accretion rates onto the black hole of between approximately
$0.0025\,M_{\odot}$/yr and $0.025\,M_{\odot}$/yr result, which are well
consistent with typically observed values for Seyfert galaxies of 
a few times $10^{-3}\,M_{\odot}$/yr to $10^{-2}\,M_{\odot}$/yr \citep{Jogee_06}.
If we assume that all of the gas transferred through the inner boundary of our
model space will reach the black hole and that the efficiency for mass energy
conversion is 10\%, this yields a source luminosity of 
$1.5 \cdot 10^{44}$erg/s corresponding to
$15\%$ of the Eddington luminosity. For radiative transfer calculations
through a simple two-dimensional axisymmetric continuous torus model,
\citet{Schartmann_05} needed a value
of $20\%$ in order to obtain a good adaptation to high spatial
resolution spectral energy data. In radiative transfer calculations, the
source luminosity sets the scaling of the reprocessed radiation. 
Determining the source luminosity is a difficult task for the case of 
NGC\,1068, being a heavily obscured Seyfert\,2 galaxy. Therefore, it can 
only be estimated with the help of the accretion disc radiation reflected 
by electrons above the dust distribution and is therefore geometry dependent. 
\citet{Pier_94} did a careful analysis of the geometry dependence and came up with
a best estimate of 

\begin{eqnarray}
\label{equ:bol_ngc1068}
L_{\mathrm{bol}} & = & 9.4 \,\cdot 10^{10} \, \left( \frac{f_{\mathrm{refl}}}{0.01} \right)^{-1} 
\left( \frac{D}{14.4\,\mathrm{Mpc}}  \right)^2 \, L_{\odot} \\
& \approx & 3.6 \cdot 10^{44}\,\frac{\mathrm{erg}}{\mathrm{s}},
\end{eqnarray} 

which corresponds to 35\% of the Eddington luminosity, including a quite large 
uncertainty\footnote{Equ.\,\ref{equ:bol_ngc1068} has been rescaled to today's 
distance estimate, compared to the original publication.}.
However, concerning the current mass accretion rate, our smooth disc 
description with an effective $\alpha$ parameter is an oversimplification. In reality, we
expect the matter transport from the circum nuclear disc towards the inner hot
accretion disc to be clumpy, as e.~g.~observed in the galactic centre
environment \citep{Montero_09}.  

\begin{figure}
\begin{center}
\includegraphics[width=0.98\linewidth]{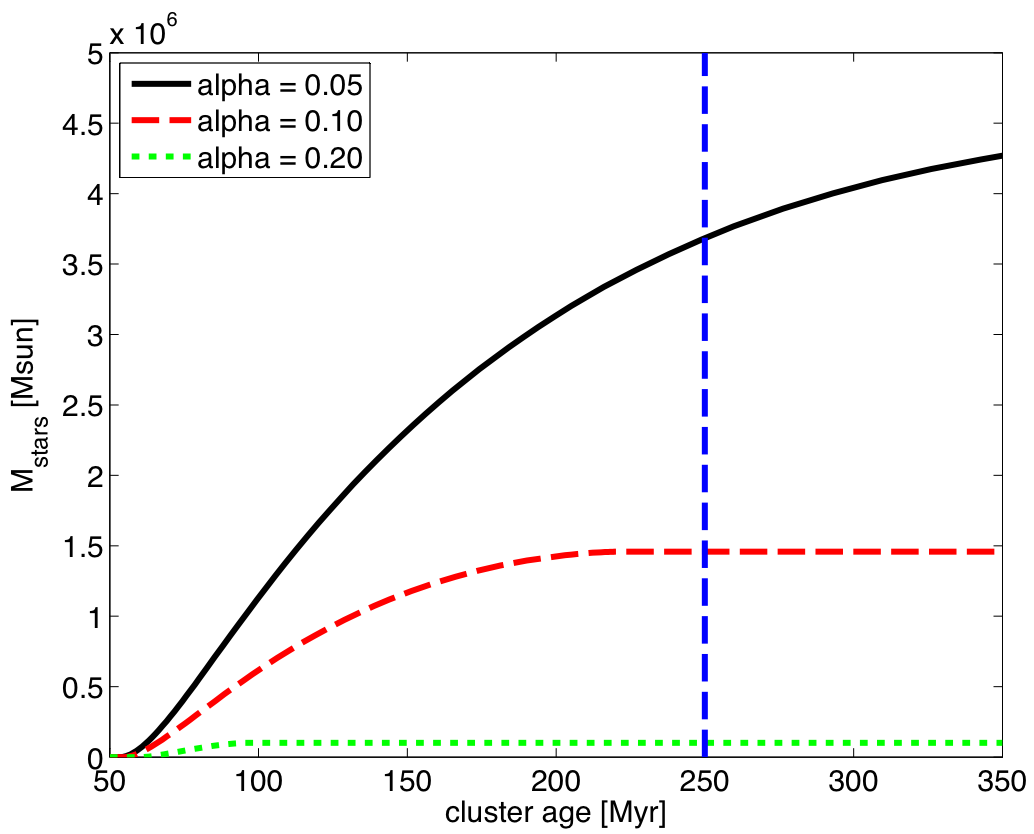}
\caption{Total mass of formed stars.}
% figure produced with:
% plot_study.m
% in mschartm@ex-sol:~/calculations/disk_evolution/paper_const_scaleheight_revised/alpha_study_kepler/
\label{fig:mstars_alpha}
\end{center}
\end{figure}

The total mass of formed stars for the $\alpha$ parameter study is shown in Fig.~\ref{fig:mstars_alpha}. 
The star formation rate increases steeply at the beginning of the simulation, when the mass input rate
is still high. Larger values of $\alpha$ lead to a larger mass accretion rate (see Fig.~\ref{fig:mdot_acc_alpha}) and a smaller
gas surface density (see Fig.~\ref{fig:gasfinal_alpha}) and in consequence to a smaller total mass in stars.  A total 
mass of up to $4\cdot 10^6\,M_{\odot}$ is reached at the current age of the nuclear starcluster. 
In comparison to this, \citet{Kumar_99} estimate the stellar mass 
within 1\,pc around the nuclear black hole in NGC\,1068 to be of the order of $10^7\,M_{\odot}$ on basis of theoretical 
considerations and from observations by \citet{Thatte_97}. Such a high rate of star formation will also lead to subsequent 
energy input into the disc by supernova type\,II explosions after a few million years. This additional feedback process
will stir further turbulence in the disc and might be an important driver to increase its scale height. However, a detailed 
analysis of these effects can only be done in three-dimensional, multi-phase hydrodynamical simulations 
including the effect of gas cooling. However, using such a scenario, 
\citet{Wada_09} find that only on scales beyond a few parsec, a significant scale height can be achieved.

A further comparison with observations can be done by calculating the neutral
hydrogen column density on the line of sight. To do this, we use
the gas surface density and distribute
the gas homogeneously in z-direction in a disc structure according to the scale height
distribution assumed previously (equation \ref{equ:scaleheight}). 
Many simplifications have to be assumed, concerning the geometrical distribution, the gas mixture,
the ionisation state, etc. Therefore, only a rough estimate can be given.
Assuming solar 
gas composition and that the disc is completely neutral, values 
between $4\times10^{25}\,\mathrm{cm}^{-2}$ for $\alpha_{\nu}=0.2$ and 
$1\times10^{26}\,\mathrm{cm}^{-2}$ for $\alpha_{\nu}=0.05$ result.
%calculated with: ex-ws40:~/work/NH_determination/calc_nh_disk_constsh.pro
In a compilation of observed Seyfert galaxies in \citet{Shi_06}, the values
spread between approximately $10^{20}\,\mathrm{cm}^{-2}$ (which is basically
the galactic foreground value) and $10^{25}\,\mathrm{cm}^{-2}$. Seyfert\,2
galaxies fill the upper part of this range starting at roughly
$10^{22}\,\mathrm{cm}^{-2}$. This shows that our derived values lie in
the upper range or even beyond the sample of Seyfert\,2 galaxies.
One should however note that a density stratification in vertical 
direction of the disc is expected. 
In particular, \citet{Mulchaey_92} find that the torus in NGC\,1068 has a very
high gas column density of the order of $N_{\mathrm{H}}= 10^{25}\,\mathrm{cm}^{-2}$. 
An additional component on tens of parsec scale, as revealed by SINFONI, adds a column density of 
approximately $8.0\times10^{24}\,\mathrm{cm}^{-2}$ \citep{Sanchez_09}.     
The hydrogen column density through our outer torus component (in our 3D hydrodynamical models) is only a negligible
fraction of these values.

%\subsubsection{Changing the amount of mass input}
%Here, we change the normalisation of the fitted mass input rate distribution
%with radius. We adapt it to the upper envelope of the real data distribution
%(see Fig.\,\ref{fig:mdot_input_histo_025}) and the lower envelope of this
%distribution. This also gives us an idea of the possible values of a clumpy
%distribution, given by the peak values within the distribution. 
   
%\subsubsection{Changing the amount of angular momentum input}  
%In this study, we change the amount of angular momentum input, by simply
%shifting the radial distribution of the mass input rate in radial direction. 

\section{Discussion}
\label{sec:discussion}

\subsection{Keeping tori stable against gravity}

One major subject in theoretical AGN torus research is the question of
stability of the vertical structure. A geometrically thick torus, which is
stabilised by Keplerian rotation will soon collapse to a thin disc, when the
thermal pressure is reduced by gas cooling or the turbulent pressure is
dissipated, e.~g.~in collisions. Being a fast process happening on the order
of only a few orbital periods, this contrasts with observations of geometrically
thick structures. A short summary of the models proposed to
circumvent this problem is given in Sect.~\ref{sec:introduction}. 

In our three-dimensional hydrodynamical model, geometrically thick gas and
dust structures naturally result 
from the fact that the emitting stars are organised in a vertically extended
structure \citep{Davies_07}. 
These clusters are {\it hot stellar systems}, which are stabilised mainly by
random motions of the stars, leading to a turbulent pressure and rotation is
only of minor importance. The turbulent motions lead only to a weak
stabilisation against gravity concerning the emitted gas, 
as the blobs merge and dissipate their random kinetic energy component 
on a short timescale and
are able to move radially inward, until they find their new equilibrium radius
in the Keplerian disc. 

However, one should note again that the outer torus part of our model can
not account for the Seyfert~1/2 dichotomy, due to its low gas column density. 
This can only be done by the inner nuclear
and very dense disc component. The crucial processes for puffing up this
structure are still a matter of debate. 
Every process which is able to stir turbulence is a possible
candidate. Turbulent motions within the disc will do both, puff-up the disc
structure to form geometrically thick tori and drive accretion as it also 
leads to an effective viscosity. 
Some ideas are discussed in Sect.~\ref{sec:turb_visc}

\subsection{Origin of the turbulent viscosity}
\label{sec:turb_visc}

Evidence is growing that magnetic fields are most important to mediate redistribution
of angular momentum in thin and ionised accretion discs. This basic idea was coined 
already by \citet{Shakura_73} after setting up their parametrised thin disc model 
({\it Shakura-Sunyaev disc, $\alpha$-disc, standard disc}). The first numerical realisation of 
this {\it Magneto-Rotational Instability} was given by \citet{Balbus_91} showing that
a weak magnetic field together with a shear flow is able to maintain a magnetic dynamo
in accretion discs and lead to an accretion flow.
This scenario still forms the basis of the most up to date simulations.
In our case, this might be relevant only for the innermost region of the disc
and a boundary layer, which can be directly ionised by the central source \citep[e.~g.~][]{Blaes_94}.
However, there is still contradiction between theoretically derived $\alpha$-values and those measured
in observations \citep{King_07}.  

Furthermore, any kind of turbulent process will lead to an effective viscosity 
and thereby drives angular momentum redistribution. Relevant for our simulations 
is first the merging of blobs and streams of gas into the nuclear disc. 
Possessing a range of various momenta coming from different directions, this 
leads to enhanced turbulence within the disc. Detailed hydrodynamical simulations 
of these processes are planned and will give us an estimate of the relevance 
for angular momentum transfer in the disc. 

\citet{Rice_09} proposed self-gravity to be the dominant transport mechanism of angular momentum
for the case of discs, which are too weakly ionised to sustain MHD turbulence \citep{Blaes_94}.
This might be the situation within some radial extent of cold and dense protoplanetary accretion discs
and might be relevant for the case of our nuclear discs as well. 
Furthermore, it has been shown that fragmentation is equivalent to an $\alpha \leq$ 0.06 
\citep{Rice_05}.

Due to the large amount of star formation
in some of our simulations, secondary feedback processes like supernova
type~II explosions might contribute significantly during some phases of the
evolution. Interaction with wind and ultra-violet radiation pressure will
play a role in the part of the  central region with direct lines of sight 
towards the nucleus.

\subsection{Dependence on the cluster rotation profile}
For the case of this paper, a simple power law is assumed for the 
radial dependence of the rotation velocity of the underlying nuclear
stellar cluster. On the tens of parsec scale, this is a fair representation
of the flat rotation curve observed with SINFONI \citep{Davies_07} for 
NGC\,1068. Given the good comparison of our results with observations as 
discussed in Sect.~\ref{sec:res_effdisc}, 
we think that our model is a possible solution. 
A different rotation velocity distribution of the stars -- as has been observed for other 
nearby Seyfert galaxies -- will lead to a change 
of the radial dependence of the mass input rate, used as source term for 
the effective disc simulations. Test calculations with several analytic input
rates showed only slightly different results. However, a detailed 
analysis of various dynamical initial conditions necessitates an 
extensive parameter study and a detailed understanding of the 
angular momentum redistribution, which is beyond the 
scope of the current paper and will be discussed in subsequent 
publications. It will finally enable us to make statistical 
statements and compare our results to 
observations of samples of galaxies.

\section{Conclusions}
\label{sec:conclusions}

In this paper, we show that evolving stars from a massive
and young nuclear star cluster, as found in nearby Seyfert galaxies provide
enough gas to assemble a parsec-sized nuclear gas disc. 
To this end, we combine an enhanced version of 
the \citet{Schartmann_09} models with a simplified treatment of the 
innermost parsec scale region, where a nuclear disc builds up.
As far as possible, we derive input parameters of our models from 
observations of the nearby and well-studied Seyfert\,2 galaxy 
NGC\,1068.
This two-stage analysis enables us to (i) do a 
long term evolution study, (ii) link the tens of parsec scale region 
of galactic nuclei (observed with the SINFONI instrument) to the sub-parsec
scales (probed by MIDI and in water maser emission) and
(iii) test our model directly with a large number of observational 
results. Such comparisons show that the total mass and surface 
density distribution of these nuclear discs are
compatible with recent observations. \newline
Our three-dimensional hydrodynamic simulations are based upon the observations of young and massive nuclear star clusters in the centres
of nearby Seyfert galaxies. In our global scenario, the gas ejection of their stars provide both, 
material for the obscuration within the so-called unified scheme
of active galactic nuclei and a reservoir to fuel the central, active region. Gas is injected in form of blobs formed by emissions of 
single intermediate mass stars with low expansion velocities. In contrast to this, turbulent velocities as taken over from the hot 
stellar system are high and contribute significantly to the clumping of the gas distribution as a whole, through interaction between
gas blobs, transforming kinetic energy into heat. Cooling instability further
confines these clouds and filaments, which then tend to move towards the minimum of the effective potential. After only a few orbital periods,
this represents a clumpy and filamentary flow of gas towards the inner region, where it opens out into a disc structure and sets into a 
stationary state in the large scale part. As we currently cannot treat the
physics of the inner part correctly in our 3D hydrodynamical simulations, we feed the mass inflow 
into our viscous disc model, in order to be able to estimate observable quantities. 
At the current age of the nuclear starburst in NGC\,1068 of 250\,yr, our simulations yield 
disc sizes of the order of 0.8 to 0.9\,pc, which compares well to the observed maser disc extent 
and disc sizes as inferred from interferometric observations in the mid-infrared. 
The gas mass of a few times $10^6\,M_{\odot}$ in the disc is in good comparison to models of discs, 
tuned to reproduce these maser observations.
The resulting mass transfer rate of $0.025\,M_{\odot}/\mathrm{yr}$ through the inner rim of our disc compares well to
typical Seyfert galaxies. However, NGC\,1068 seems to be in a heavily accreting state, which can be accomodated in our
model only, when assuming e.\,g.\,clumpy accretion. 
Furthermore, rough estimates of gas column densities of our model are in agreement with NGC\,1068 being a Compton thick source. 
On basis of these comparisons, we conclude that the proposed scenario seems to be a reasonable model and shows that 
nuclear star formation and subsequent AGN activity are intimately related.

%\subsection{Used tools for the effective disc model}   

%\begin{dingautolist}{172}
%\item the needed input from the 3D PLUTO simulations is generated with the
%  help of the following tools (to be found in
%  mschartm@ex-srv4:~/data/routines/IDL/PLUTO\_IDL): \\
  
%  accretionratesnap.pro: calculate the current and average accretion rate
%  through a sphere with specified radius \\

%  disk\_matlab\_input.pro: calculate histograms of e.g. mass input vs. radius of
%  input, to which an analytical expression is fitted!

%\item the one-dimensional model for the nuclear disc is calculated with the
%  help of the following matlab routine, matlab's pdepe function is used to
%  solve the partial differential equation: \\
%  mschartm@ex-sol:~/calculations/disk\_evolution 
%\end{dingautolist}

\section*{Acknowledgments}
We thank the referee, Keichi Wada, for useful comments to improve the 
clarity of the paper and Volker Gaibler for helpful discussions.
Part of the numerical simulations have been carried out on the SGI 
Altix 4700 HLRB\,II of the Leibniz Computing Centre in Munich (Germany). 

\bibliographystyle{mn2e}
\bibliography{astrings,literature}

\end{document}